\documentclass[a4paper,11pt]{article}
\usepackage{pos}

\usepackage{csquotes}    
\usepackage{tikz}        
\usepackage{braket}      
\usepackage{placeins}    
\usepackage{arydshln}    

\newcommand{\overlayTextBox}[6]{
  \begin{tikzpicture}
    \centering
    \node at (0, 0) {\includegraphics[width=#1,keepaspectratio,origin=c,page=#6]{#2}};
    \node[draw,rounded corners] at (#3,#4) {#5};
  \end{tikzpicture}
}

\DeclareRobustCommand{\order}[1]{\mathcal{O}\left(#1\right)}

\newcommand{\tcr}[1]{\textcolor{black}{#1}}

\newcommand{\rb}[1]{\!\left(#1\right)}
\newcommand{\sq}[1]{\left[#1\right]}
\newcommand{\expe}[1]{\text{e}^{#1}}
\DeclareRobustCommand{\eqnr}[1]{Eq.~$\left(\ref{#1}\right)$}

\newcommand{\Fig}[1]{Fig.~\ref{#1}}
\newcommand{\Tab}[1]{Table \ref{#1}}

\newcommand{\Refl}[1]{Ref.~\cite{#1}}    
\newcommand{\Refltwo}[2]{Refs.~\cite{#1,#2}}    
\newcommand{\Reflthree}[3]{Refs.~\cite{#1,#2,#3}}    

\title{Finite temperature hadronic spectral properties}

\author*[a]{Ryan Bignell}
\author[b]{Gert Aarts}
\author[b]{Chris Allton}
\author[b]{M.~Naeem Anwar}
\author[b]{Timothy J. Burns}
\author[c]{Rachel Horohan D'arcy}
\author[d]{Benjamin J\"ager}
\author[e]{Seyong Kim}
\author[f]{Maria Paola Lombardo}
\author[a]{Sin\'ead Ryan}
\author[a,c]{Jon-Ivar Skullerud}
\author[b]{Antonio Smecca}

\affiliation[a]{School of Mathematics \& Hamilton Mathematics Institute,
  Trinity College, Dublin, Ireland}

\affiliation[b]{Department of Physics,
   Swansea University, Swansea, SA2 8PP, United Kingdom}

\affiliation[c]{Department of Physics
  National University of Ireland Maynooth, County Kildare, Ireland}

\affiliation[d]{CP3-Origins \& Danish IAS, Department of Mathematics and Computer Science,\\
  University of Southern Denmark, 5230, Odense M, Denmark}
\affiliation[e]{Department of Physics,
  Sejong University, Seoul 143-747, Korea}

\affiliation[f]{INFN,
  Sezione di Firenze, 50019 Sesto Fiorentino (FI), Italy}

\emailAdd{bignellr@tcd.ie}
\emailAdd{\{g.aarts,c.allton,m.n.anwar,t.burns,antonio.smecca\}@swansea.ac.uk}
\emailAdd{rachel.horohandarcy.2018@mumail.ie}
\emailAdd{jaeger@imada.sdu.dk}
\emailAdd{skim@sejong.ac.kr}
\emailAdd{mariapaola.lombardo@lnf.infn.it}
\emailAdd{ryan@maths.tcd.ie}
\emailAdd{jonivar@thphys.nuim.ie}

\abstract{The FASTSUM collaboration has a long-standing project examining hadronic properties using anisotropic lattice QCD. We determine the spectral properties of bottomonia at finite temperature using lattice NRQCD and describe how our newer simulations improve our control over systematic errors. Motivated by these efforts, the temperature dependence of charm hadron masses is determined where it is found that temperature effects can extend into the confining phase and that some species remain stable deep past the pseudo-critical temperature.}

\FullConference{The XVIth Quark Confinement and the Hadron Spectrum Conference (QCHSC24)\\
 19-24 August, 2024\\
 Cairns Convention Centre, Cairns, Queensland, Australia\\}

\begin{document}
\maketitle
\section{Introduction}
Hadrons containing heavy quarks are important probes of the dynamics of the Quark Gluon Plasma (QGP) as they are relatively insensitive to the dynamics of the chiral transition. Modification to these hadrons may hence be described by potential models. In particular, these dynamics are encoded in the spectral function $\rho\rb{\omega}$.
\par
The \textsc{Fastsum} collaboration has a long standing program~\cite{Aarts:2010ek,Aarts:2011sm,Aarts:2014cda,JonivarProceedings} of the computation of bottomonium spectral functions in the quark gluon plasma using lattice NRQCD~\cite{Lepage:1992tx}. Here we provide a comparison between two of our thermal ensembles Generation 2 and Generation 2L which differ only in their (non-physical) pion mass and hence provide insight into whether this systematic is under control.
\par
As an extension of our previous work on light and strange baryons~\cite{Aarts:2015mma,Aarts:2017rrl,Aarts:2018glk}, we also examine the behaviour of spin $1/2$ charm baryons as a function of temperature~\cite{Aarts:2023nax}. As the mass of the charm quark is much heavier than that of the light or strange, one expects a reduced degree of chiral symmetry restoration in the QGP. We report here on our determination of the ground state positive and negative parity masses as a function of temperature.
\section{Ensembles}
\label{sec:LQCD:Ens}
\begin{table}[b]
  \centering
    \caption{\textsc{Fastsum} Generation 2 and 2L ensembles used in this work. The solid vertical lines indicates $T_c^{\text{Gen2L}}=167(2)(1)$ MeV~\cite{Aarts:2020vyb,Aarts:2022krz} and the dashed for Generation 2 where $T_c^{\text{Gen2}}=181(1)$ MeV~\cite{Aarts:2019hrg}. $T_c$ is determined using the renormalised chiral condensate. Full details of these ensembles may be found in Refs.~\cite{Aarts:2020vyb,Aarts:2022krz}.
    }
  \begin{tabular}{r|r|rrrrr||rr;{1pt/1pt};{1pt/1pt}rrrr}
    &$N_\tau$ & 128 & 64 & 56 & 48 & 40 & 36 & 32 & 28 & 24 & 20 & 16\\ \hline
    Gen 2 &$T\,\rb{\text{MeV}}$ & 44 &  &  & & 140 & 156 & 176 & 201 & 235 & 281 & 352 \\
    Gen 2L &$T\,\rb{\text{MeV}}$ & 47 & 95 & 109 & 127 & 152 & 169 & 190 & 217 & 253 & 304 & 380 \\
\end{tabular}
\label{tab:ensembles}
\end{table}
The thermal ensembles of the \textsc{Fastsum} collaboration~\cite{Aarts:2014nba,Aarts:2020vyb,aarts_2023_8403827,Aarts:2022krz,aarts_2024_10636046} are used in this study. These ensembles use the \enquote{fixed-scale} approach wherein the temperature, $T=1/\rb{a_\tau N_\tau}$ is varied by changing the temporal extent $N_\tau$. This, in conjunction with an anisotropic approach where the temporal lattice spacing $a_\tau$ is much smaller than the spatial one $a_s$, allows for a finely spaced spread of temperatures.
\par
We utilise the Symanzik-improved anisotropic gauge and mean-field-improved Wilson-clover action of the Hadron Spectrum Collaboration~\cite{Edwards:2008ja}. Full details of the action and parameter values can be found in \Refl{Aarts:2020vyb}. For Generation 2, $a_s = 0.1205\rb{8}$ while Generation 2L has $a_s = 0.11208\rb{31}$ fm~\cite{Wilson:2019wfr} with pion masses  of $m_\pi=384(4), 239(1)$ MeV respectively. With anisotropies $\xi=a_s/ a_t = 3.444(6)$ and $\xi=3.453(6)$ for Generation 2 and 2L respectively, these ensembles have nearly identical parameters aside from the pion mass. For the NRQCD on Generation 2L we use the (earlier) scale setting from \Refl{Wilson:2015dqa}, $a_\tau=0.0330(2)$ fm, in this study. The strange quark has been approximately tuned to its physical value via the tuning of the light and strange pseudoscalar masses~\cite{HadronSpectrum:2008xlg,HadronSpectrum:2012gic,Cheung:2016bym}.
\par
The lattice size for Generation 2 is $24^3 \times N_\tau$ while it is $32^3 \times N_\tau$ for Generation 2L, with temperature $T = 1/\rb{a_\tau N_\tau}$. For Generation 2L we use $\sim 1000$ configurations with up to $N_\tau$ Coulomb gauge fixed wall-sources for the NRQCD correlators. The relativistic baryon correlators have eight sources each. Generation 2 has $\sim 500$ configurations at the two lowest temperature and $\sim 1000$ at all others. A summary of the ensembles is given in \Tab{tab:ensembles}.
\section{Bottomonia}
The spectral properties of bottomonium states are encoded in the correlation function
\begin{align}
  G\rb{\tau; T} \int_{\omega_{\text{min}}}^{\infty}\,\frac{d\,\omega}{2\,\pi}\,K\rb{\tau,\omega; T}\,\rho\rb{\omega;T},
  \label{eqn:Grho}
\end{align}
where $K\rb{\tau,\omega;T}$ is the known kernel function and $\rho\rb{\omega;T}$ is the spectral function at temperature $T$. In the relevant case of non-relativistic quantum chromodynamics (NRQCD), this kernel is the same at all temperatures and is given by
\begin{align}
  K\rb{\tau,\omega} = \expe{-\omega\,\tau}.
\end{align}
The spectral properties are important to the analysis of the quark gluon plasma produced in heavy-ion collision experiments where bottomonium states are created in the initial hard collisions and so may carry information about these early stages.
\par
Given a finite number of data points, the reconstruction of the spectral function $\rho\rb{\omega; T}$ from the (Euclidean) correlator $G\rb{\tau; T}$ is a numerically ill-posed problem \tcr{(see i.e. \Refl{Jay:2025dzl} and references therein)} requiring regularisation. A common subset of approaches are Bayesian, and in particular we make use of the Maximum-Entropy-Method (MEM)~\cite{Asakawa:2000tr,Aarts:2007pk,Aarts:2011sm}.
\subsection{Maximum-Entropy-Method}
The Maximum-Entropy-Method regularises the ill-posed problem through the inclusion of additional information in the form of the prior (or equivalently the default model $m\rb{\omega}$). This information is incorporated using Bayes theorem which may be written
\begin{align}
  P\rb{\rho|DI} = \frac{P\rb{D|\rho I}\,P\rb{\rho|I}}{P\rb{D|I}},
\end{align}
where $D$ is the data and we parameterise the prior probability by
\begin{align}
  P\rb{\rho|I} \propto \expe{\alpha\,S\sq{\rho}} \implies P\rb{\rho|DI}\propto\expe{-L\sq{D,\rho} + \alpha\,S\sq{\rho}}.
\end{align}
Here $L$ is the standard $\chi^2$ likelihood and $S\sq{\rho}$ is the Shannon-Jaynes entropy~\cite{Asakawa:2000tr}
\begin{align}
  S = \int\,d\omega\,\rho\rb{\omega}\,\log\frac{\rho{\omega}}{m\rb{\omega}}.
\end{align}
The spectral function is then expressed in terms of the default model $m\rb{\omega}$
\begin{align}
  \rho\rb{\omega} = m\rb{\omega}\,\exp\rb{\sum_{k=1}^{N_b}\,b_k\,u_k\rb{\omega}},
\end{align}
where $b_k$, $u_k$ are found via singular-value-decomposition and Bryan's method~\cite{bryan_maximum_1990,Asakawa:2000tr} (see \Refl{Aarts:2011sm} for details of our implementation). A notable addition however is the use of the Laplace shift technique introduced in \Refltwo{Page:2021ohe}{BenPageThesis} at zero temperature $T_0$ (corresponding to $N_\tau=128$) for Generation 2L. Here the correlator is \enquote{shifted} by a factor $\expe{\Delta\,\tau}$ before applying the MEM to obtain the shifted spectrum $\rho\rb{\omega - \Delta}$. This brings the system closer to small $\omega$ where the resolving power of the method is greater. The factor $\Delta$ can then be easily removed to obtain the unshifted spectrum $\rho\rb{\omega}$.
\begin{figure}[p!]  
  \centering
    \begin{tikzpicture}
    \centering
    \node at (0, 0) {\includegraphics[width=0.48\columnwidth, keepaspectratio,origin=c,page=3]{Figures/grExportPlot_128.pdf}};
    \node at (-1.5, 1.3) {\tiny$\Upsilon\rb{1S}$};
    \node at (-0.4, 0.6) {\tiny$\Upsilon\rb{2S}$};
    \node at (0.25, 0.0) {\tiny$\Upsilon\rb{3S}$};;
    \node at (0.75, 0.6) {\tiny$\Upsilon\rb{4S}$};;

    \node[draw, rounded corners] at (2.2, 0.95) {\tiny$T\sim47$ MeV};;
    \end{tikzpicture}
    \overlayTextBox{0.47\columnwidth}{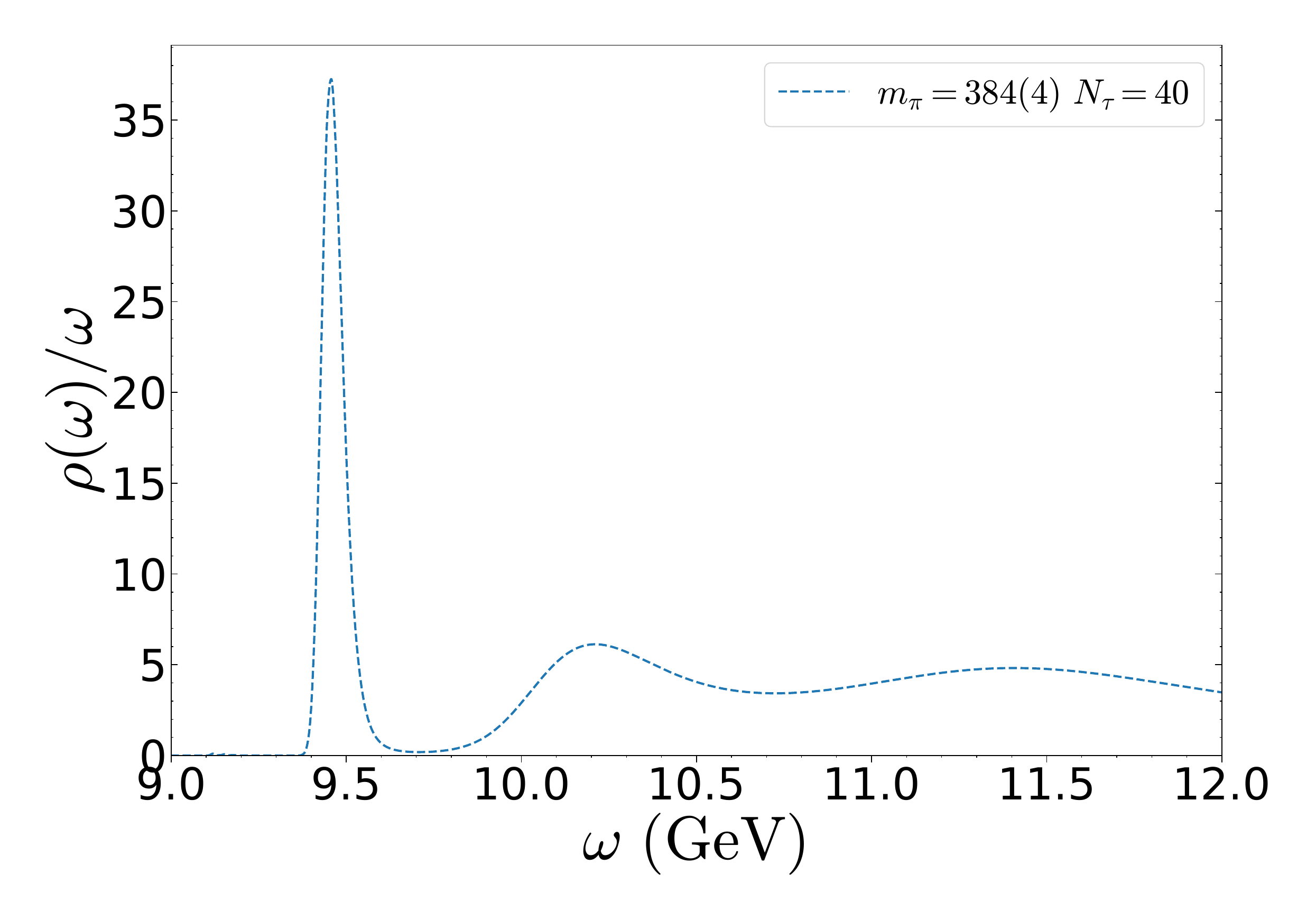}{2.2}{0.95}{\tiny$T\sim150$ MeV}{2}
    \\
    \overlayTextBox{0.47\columnwidth}{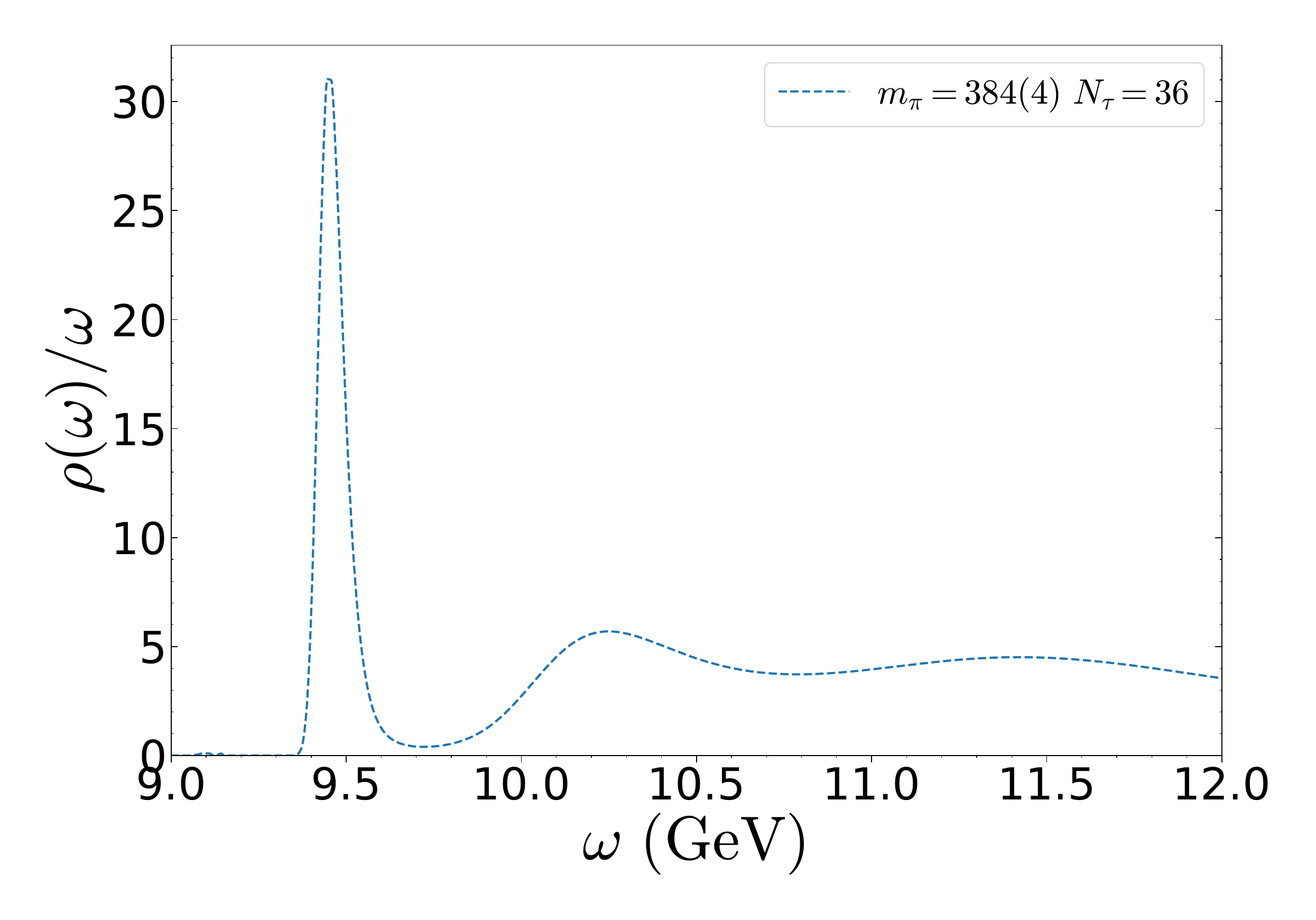}{2.2}{0.95}{\tiny$T\sim165$ MeV}{2}
    \overlayTextBox{0.47\columnwidth}{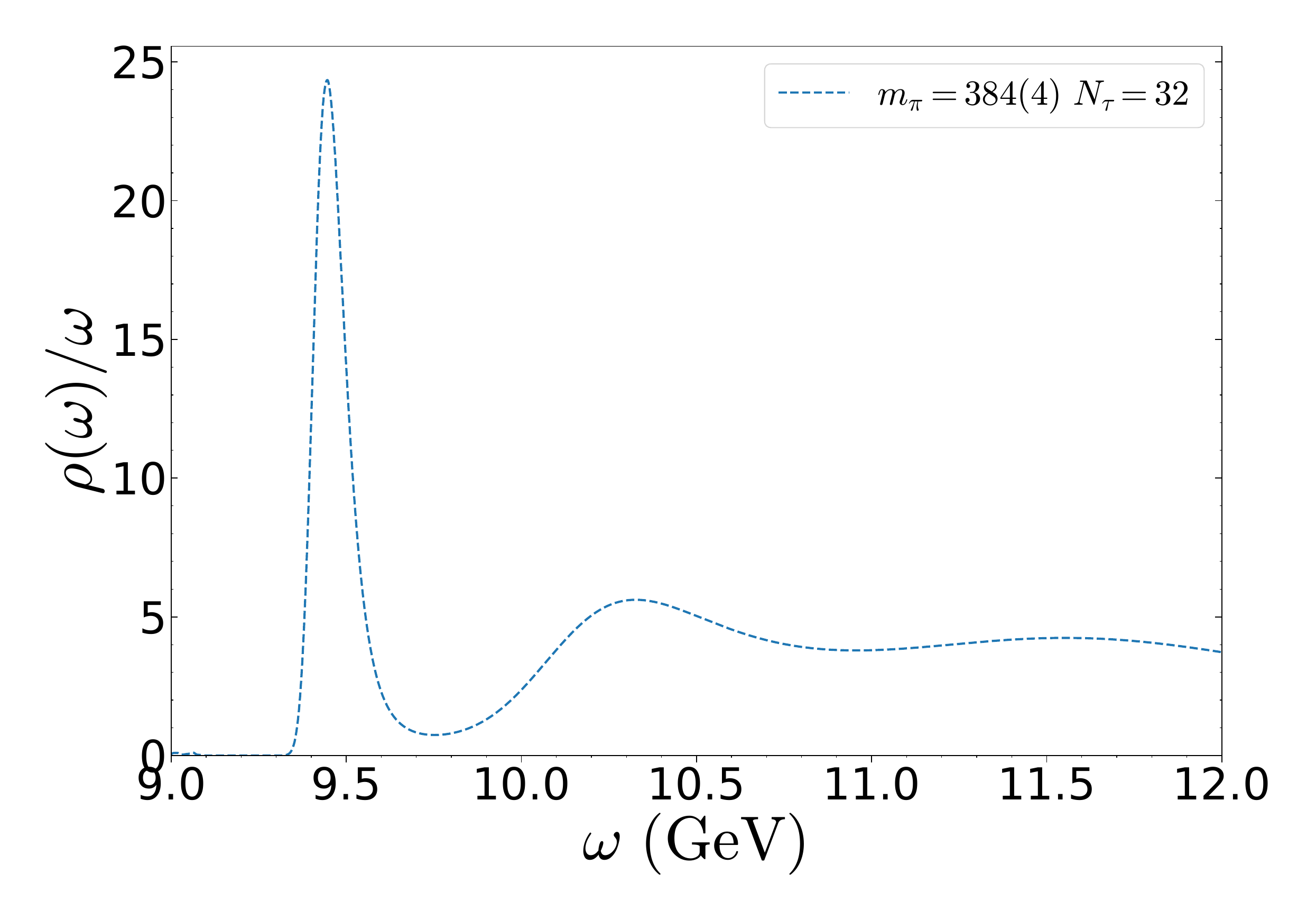}{2.2}{0.95}{\tiny$T\sim185$ MeV}{2}
    \\
    \overlayTextBox{0.47\columnwidth}{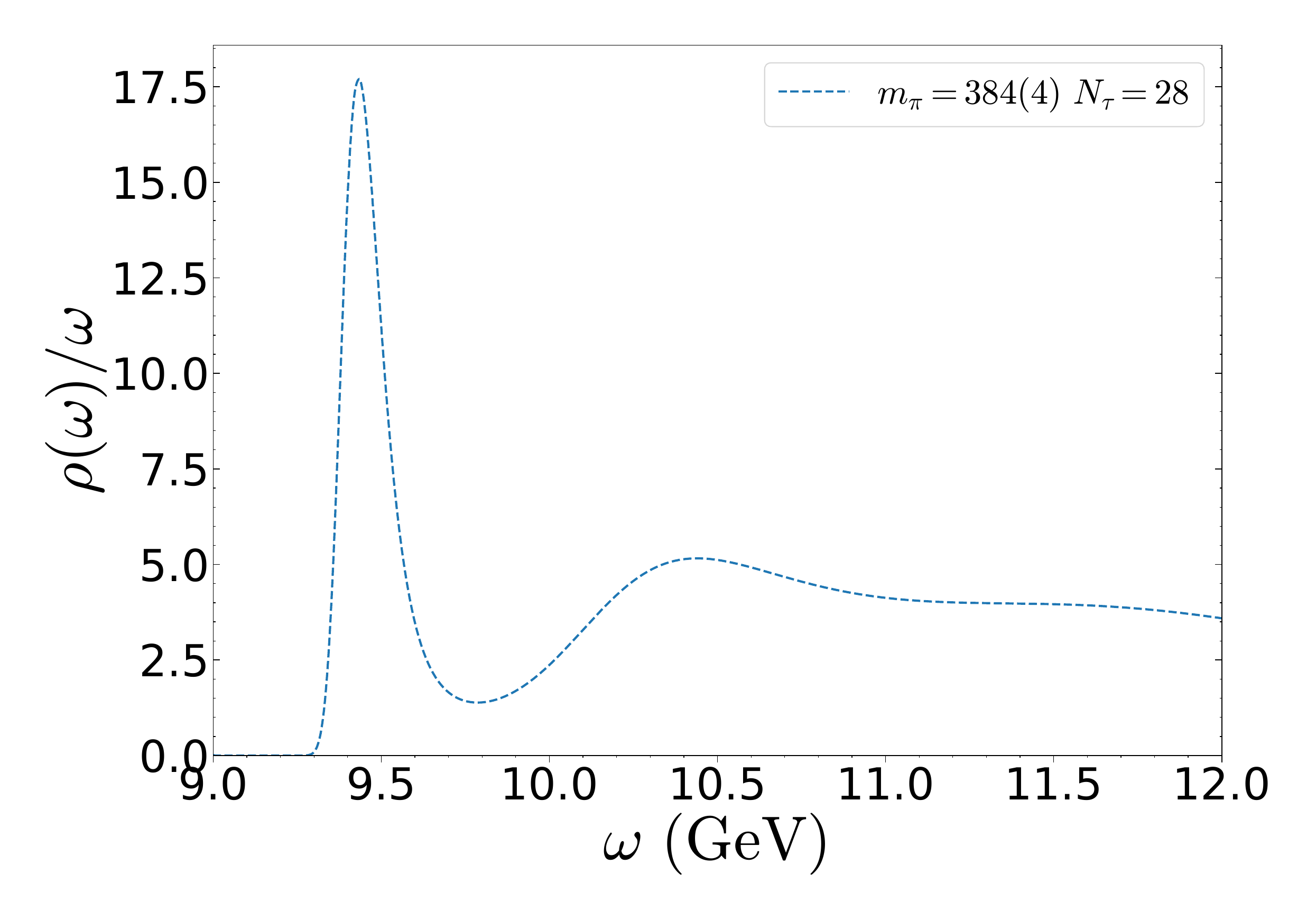}{2.2}{0.95}{\tiny$T\sim210$ MeV}{2}
    \overlayTextBox{0.47\columnwidth}{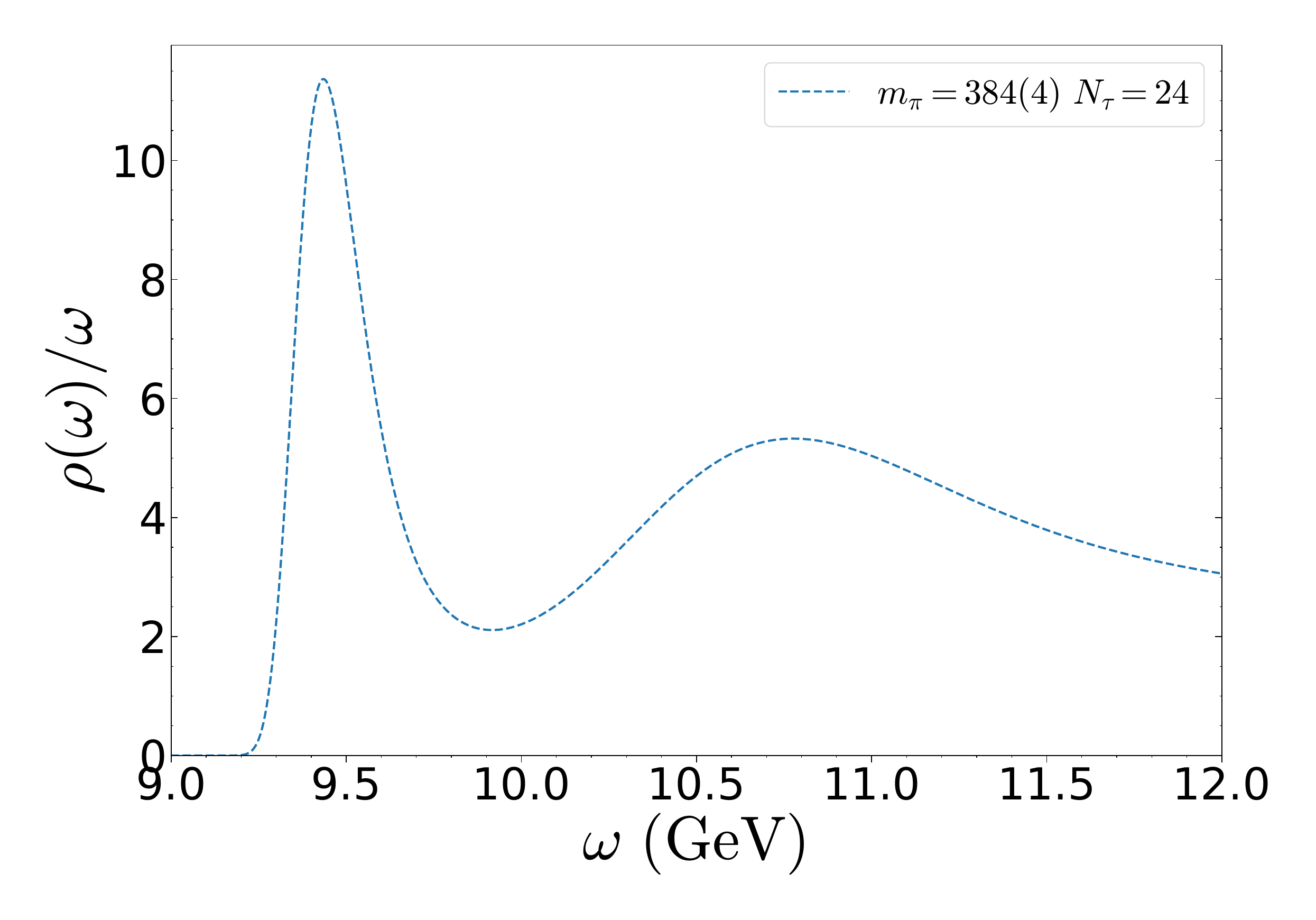}{2.2}{0.95}{\tiny$T\sim245$ MeV}{2}
    \\
    \overlayTextBox{0.47\columnwidth}{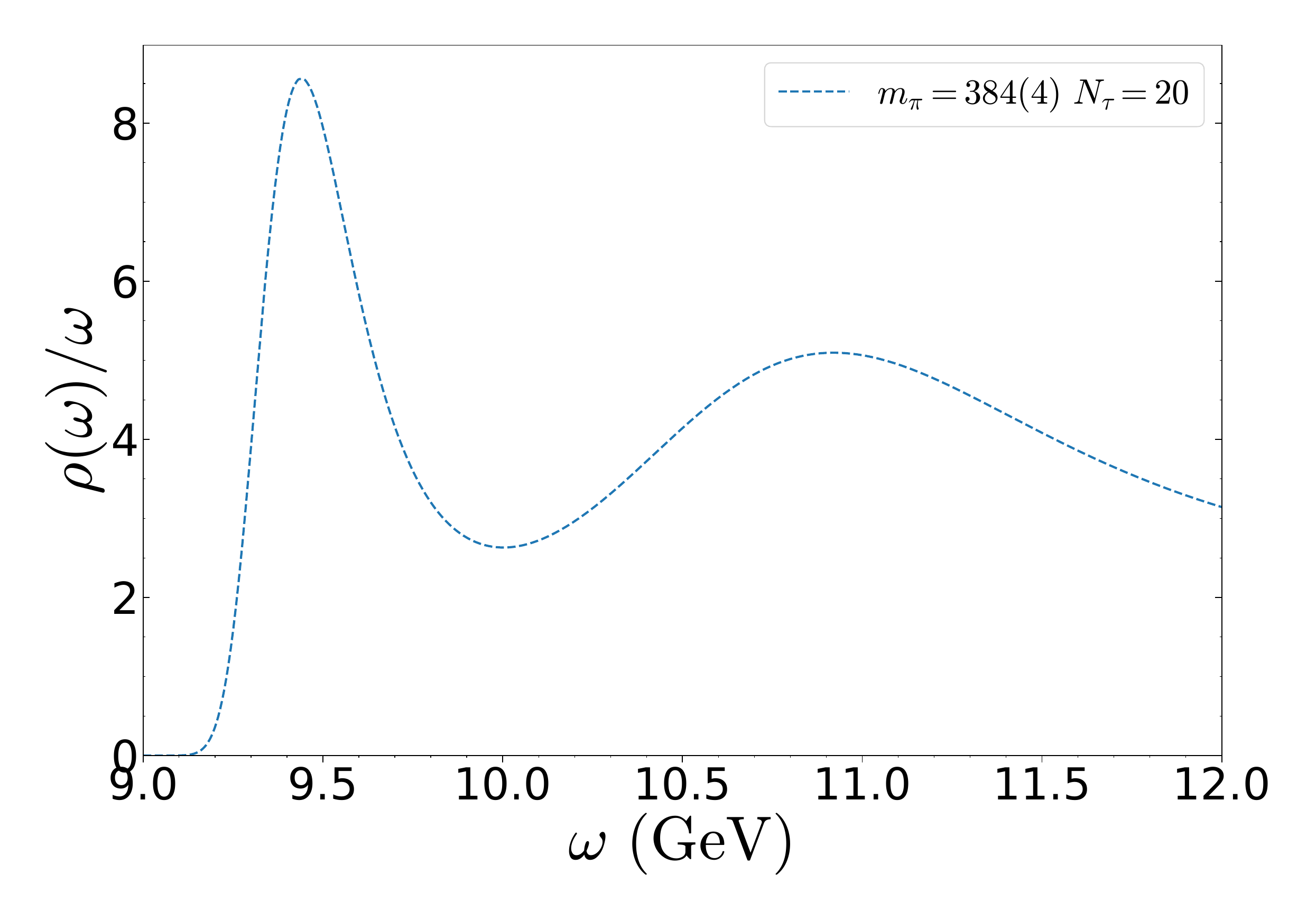}{2.2}{0.95}{\tiny$T\sim295$ MeV}{2}
    \overlayTextBox{0.47\columnwidth}{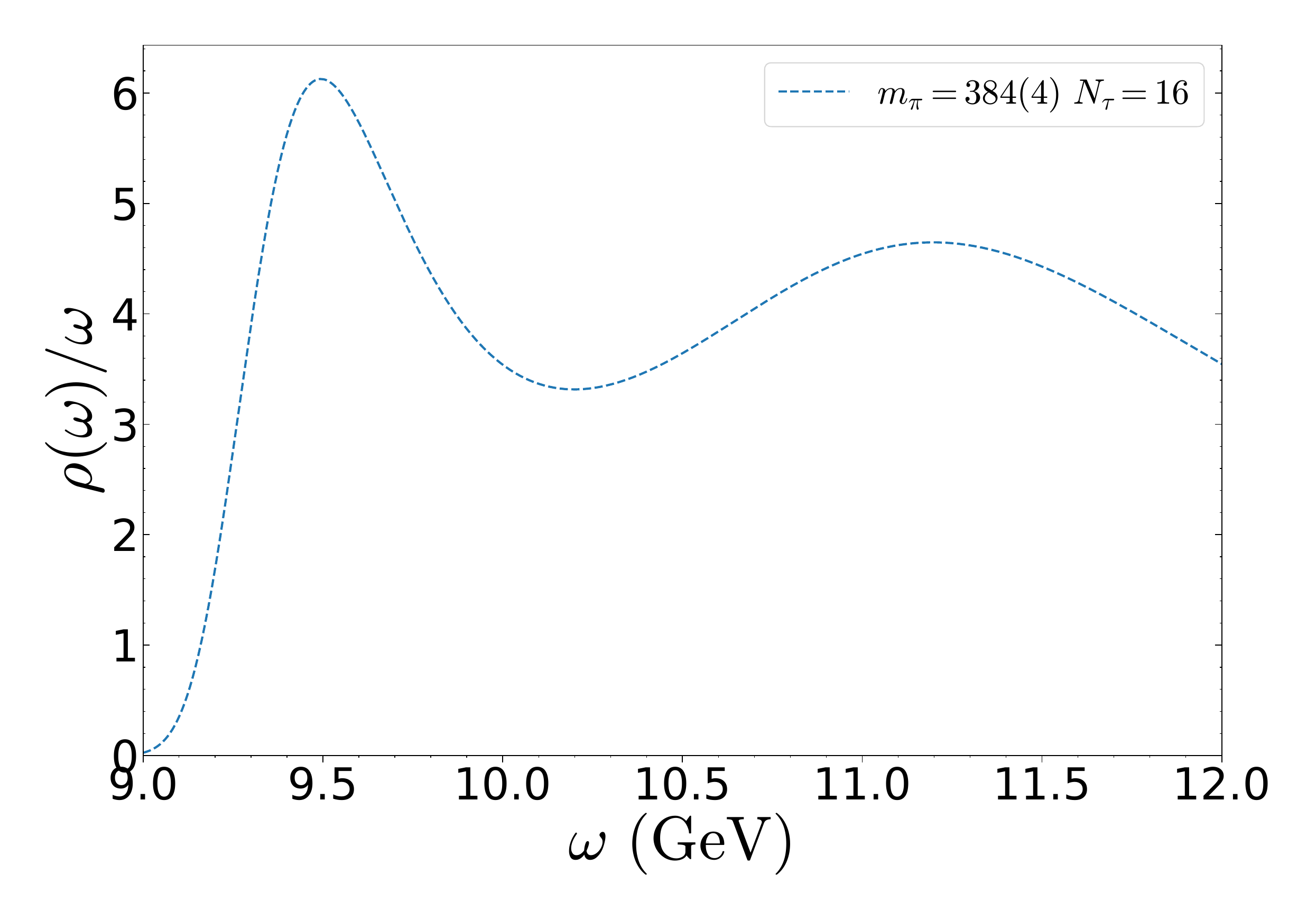}{2.2}{0.35}{\tiny$T\sim365$ MeV}{2}
  \caption{$\Upsilon$--$\,$channel MEM spectral functions for the common temperatures between Generation 2 and Generation 2L. At zero temperature $\rb{N_\tau=128}$ the experimental results from the Particle Data Group~\cite{PhysRevD.110.030001} are shown. Note the $x$-axis is common between each plot, but that the $y$-axis differs.}
  \label{fig:MEM_G22L}
\end{figure}
\par
In our simulations of bottomonia, the bottom-quark mass has been tuned through exponential fits to a point-point correlator such that the spin-averaged (1S) state has a mass which is consistent with experiment. However the light -- up \& down -- quarks in the sea are heavier than physical though Generation 2L is much closer to the physical point. In \Fig{fig:MEM_G22L} we show a comparison of the $\Upsilon$ spectral functions obtained using the MEM at all common temporal extents, $N_\tau$. The most obvious difference is in the top left figure, where the spectral function at zero temperature is shown. Here the peaks corresponding to the $\Upsilon(1S)$ and the $\Upsilon(2S)$ are closer closer to their experimental values for Generation 2L than for Generation 2. The shift for the $\Upsilon(2S)$ cannot be accommodated by the uncertainty in the lattice spacing and the mass tuning, and so most likely represents a physical effect of the lighter sea quark mass. \tcr{Although there is a peak at around the $\Upsilon(4S)$ mass at zero temperature, the absence of the $3S$ state is unexplained and so we do take this to be the $4S$ state.}
\par
The same shift of Generation 2L to smaller masses also appears at finite temperature and appears to be increasing as the temperature increases. This change is smaller than the uncertainties due to scale setting and mass tuning. The excited states display less peaked behaviour as the temperature increases with the structure at around the mass of the $\Upsilon(4S)$ immediately not discernible and the $\Upsilon(2S)$ increasingly broad. Due to the width of these states, it is unlikely that any sea mass dependence can be extracted.
\subsection{Generalised eigenvalue problem}
An improved method to extract ground state properties at zero temperature is formed by considering a symmetric matrix of operators instead of a single operator. The matrix can be diagonalised by solving a generalised eigenvalue problem~\cite{Michael:1985ne,Blossier:2009kd} (GEVP) such that a new set of operators, each coupling only to a single state, is formed. In these proceedings, we make use of the recently discussed results of \Refl{Bignell:2025bga} for the ground state $\Upsilon(1S)$ as a function of temperature.
\par
\begin{figure}[t!]
  \centering
  \includegraphics[width=0.49\columnwidth]{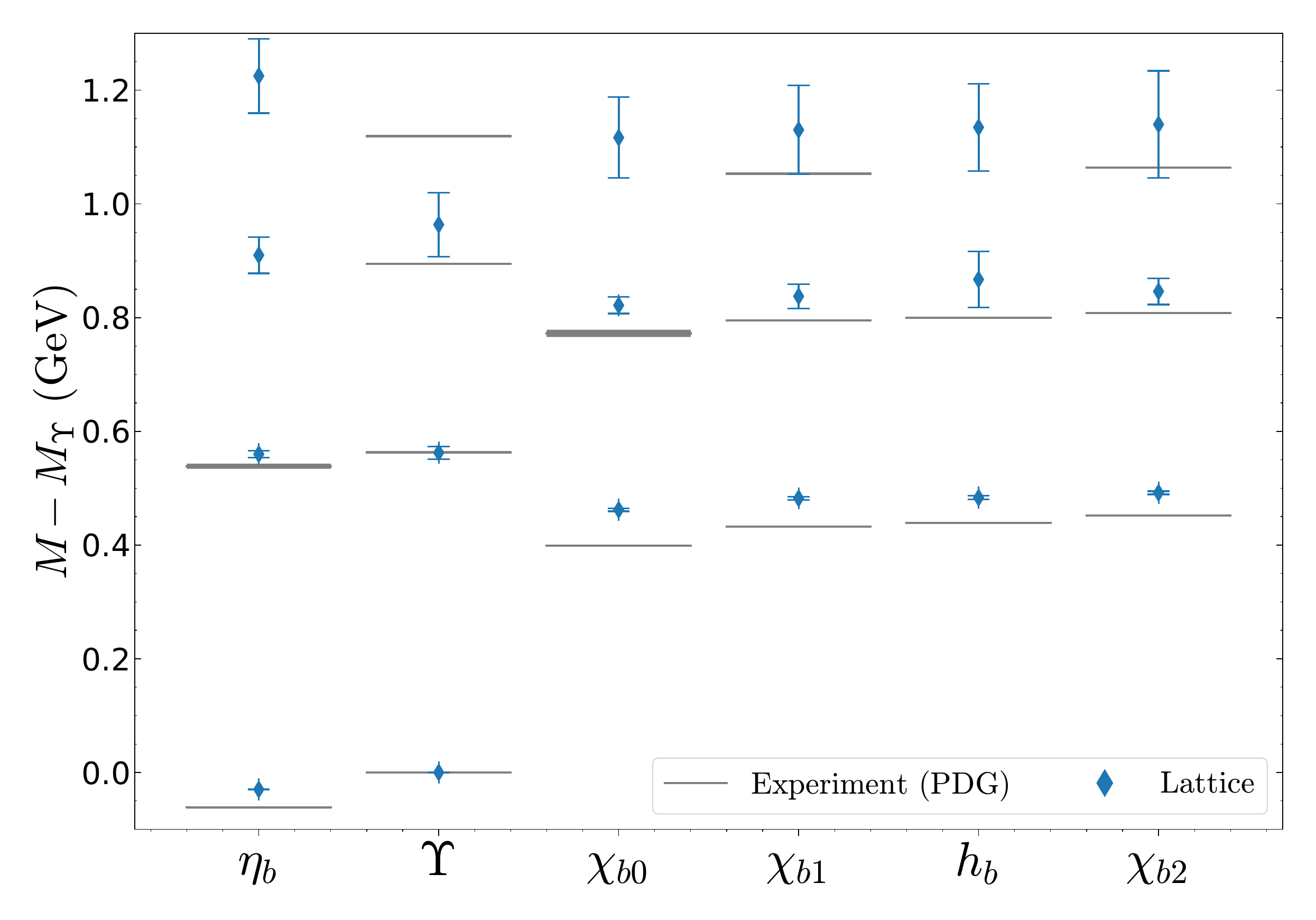}
  \includegraphics[width=0.49\columnwidth,page=2]{Figures/spect_combined_Lorentzian.pdf}
  \caption{\textbf{Left:} zero-temperature bottomonium masses obtained using the GEVP method. The $\Upsilon(1S)$ mass has been subtracted off in each case as it is used to set the NRQCD additive mass shift. The experimental results are from the Particle Data Group~\cite{PhysRevD.110.030001}. \textbf{Right:} zero temperature $\rb{N_\tau = 128}$ subtracted $\Upsilon(1S)$ mass as a function of temperature using the MEM approach compared with results using standard (multi-)exponential fits. The zero-temperature mass is determined separately for each analysis, i.e. $i=\text{GEVP}$ or $i=\text{MEM}$.}
  \label{fig:GEVP}
\end{figure}

Here we used standard multi-exponential fits of the form
\begin{align}
  G_{\text{fit}} = \sum_{i=1}^{N_{\text{exp}}}\,A_i\,\expe{-E_i\,\tau},
  \label{eqn:barfit}
\end{align}
with $N_{\text{exp}} \in \sq{1,9}$. These fits were performed at all temperatures for which the ratio of the zero and non-zero-temperature correlators suggest that the temperature effects are mild and thus an exponential fit analysis remains appropriate. A model averaging approach~\cite{Rinaldi:2019thf,NPLQCD:2020ozd,Jay:2020jkz} aids a robust determination of the mass. The zero-temperature results are shown in \Fig{fig:GEVP} (left) where the $\Upsilon(1S)$ mass is subtracted off as it is used to set the additive NRQCD mass shift. Here the $2S$ and $3S$ S-wave $\Upsilon$ states reproduce their experimental values, but the P-wave states $\text{ }\!\rb{\chi_{b0},\,\chi_{b1},\,\chi_{b2}, h_b}$ and the $\eta_{b}$ are systematically heavier. This is because we include only terms up to $\order{v^4}$ in the NRQCD expansion with tree-level coefficients~\cite{HPQCD:2011qwj,Hudspith:2023loy}.
\par
The absence of the $\Upsilon(3S)$ from the MEM spectrum is curious as it is reproduced by the GEVP analysis. Note that the MEM analysis uses a point-point correlator whereas the GEVP analysis uses a matrix of correlators which does not include the point-point correlator -- and so it would be interesting to apply the MEM analysis to GEVP projected correlators.
\par
From the ratio analysis of \Refl{Bignell:2025bga} showing mild thermal effects in the correlators up to $T=253$ MeV, for the $\Upsilon(1S)$, we fit all temperatures up to (including) this temperature. This is in accord with other work reviewed in \Reflthree{Andronic:2024oxz}{Massacrier:2024odp}{Andronic:2025jbp} which suggests that the $\Upsilon(1S)$ remains bound well past the chiral transition temperature, $T_c$.
\subsection{Mass Comparison}
We can now compare the (zero-temperature subtracted) mass with the MEM results from the previous section. This is shown in \Fig{fig:GEVP} (right). The MEM and exponential fit results are remarkably similar and suggest that there is a small, negative shift in the mass as the temperature increases. \tcr{The increased mass seen in the MEM results after $T=250$ MeV may be an analysis artefact caused by the significant reduction in the number of data points $\rb{N_\tau}$ available~\tcr{\cite{JonivarProceedings}}. In this light, the MEM results are may be an underestimate of the true negative mass shift}. Similarly, if there is residual excited state contamination, the GEVP exponential fit results would be an underestimate of the true negative mass shift.
\par
The qualitative agreement between the MEM and exponential fit approaches is remarkable as they are very different methods. The exponential fit relies upon modelling the correlator as a series of $\delta$-functions while the Bayesian MEM approach allows significant width in the spectral function, visible in the spectral functions of \Fig{fig:MEM_G22L}. We can draw significant confidence in the observed small negative mass shift trend as the values are in agreement at some temperatures and close at the rest. However no such claim about the width can be made -- the exponential fits can tell us nothing about the width of the state. Recent progress by \textsc{Fastsum} on alternative methods to determine the width is presented in \Reflthree{Darcy:2025tzj}{Smecca:2025hfw}{JonivarProceedings}.
\section{Charm Baryons}
We turn our attention to charm baryons in the spin $1/2$ sector using the Generation 2L ensembles. Doubly charmed baryons may behave in a similar manner to heavy-quarkonia and hence persist past the chiral transition temperature $T_c$. In this work~\cite{Aarts:2023nax}, we use an approach similar to that of the NRQCD exponential fits discussed above wherein a ratio of correlators at different temperatures is examined to determine where standard exponential fits are performed. The differences for relativistic baryons will be summarised.
\par
The appropriate fit function for a relativistic baryon is
\begin{align}
  \overline{G}\rb{\tau} = \sum_{n=1}^{N_{\text{exp}}}\,A_n\,\expe{-a_\tau\,M_n^+\,\tau/a_\tau} + B_n\,\expe{-a_\tau\,M_n^{-}\,\rb{N_\tau - \tau / a_\tau}},
  \label{eqn:GBaryon}
\end{align}
where $M_n^{\pm}$ is the $n^{\text{th}}$ positive (negative) parity state and the correlator $\overline{G}\rb{\tau} = G_{+}\rb{\tau} / G_{+}\rb{0}$ has been normalised at the source location. We allow the number of states $N_{\text{exp}}$ to vary from $1$ to $3$ and use model averaging techniques~\cite{Jay:2020jkz,NPLQCD:2020ozd,Rinaldi:2019thf} to ensure a robust determination of the mass. Full details of this procedure may be found in \Refl{Aarts:2023nax}.
\par
The Euclidean correlator of a baryon may be written as in \eqnr{eqn:Grho}, but now the known kernel function takes the form
\begin{align}
  K_{F}\rb{\tau, \omega; T} = \frac{\expe{-\omega\,\tau}}{1 + \expe{-\omega / T}}.
\end{align}
As this kernel has explicit temperature dependence in order to compare spectral functions $\rho\rb{\omega;T}$, at different temperatures, we can not simply take the ratio of the zero and non-zero-temperature correlators as for the NRQCD case. Instead we construct a model correlator $G_{\text{model}}\rb{\tau; T, T_0}$ at each temperature using the ground state parameters determined at zero temperature, $T_0$, using \eqnr{eqn:GBaryon} and take a ratio with that. To further elucidate the effect of temperature, we consider a \enquote{double} ratio
\begin{align}
  R\rb{\tau;T,T_0} & = \left.\frac{G\rb{\tau; T}}{G_{\rm model}\rb{\tau; T, T_0}}\right/\frac{G\rb{\tau; T_0}}{G_{\rm model}\rb{\tau; T_0, T_0}},
\end{align}
which additionally acts to eliminate the effects of excited states such that a ratio close to one is a sign of minimal changes to the spectral function. Note that in the NRQCD case, the model correlators would exactly cancel. 
\begin{figure}[t]
  \includegraphics[width=0.5\columnwidth]{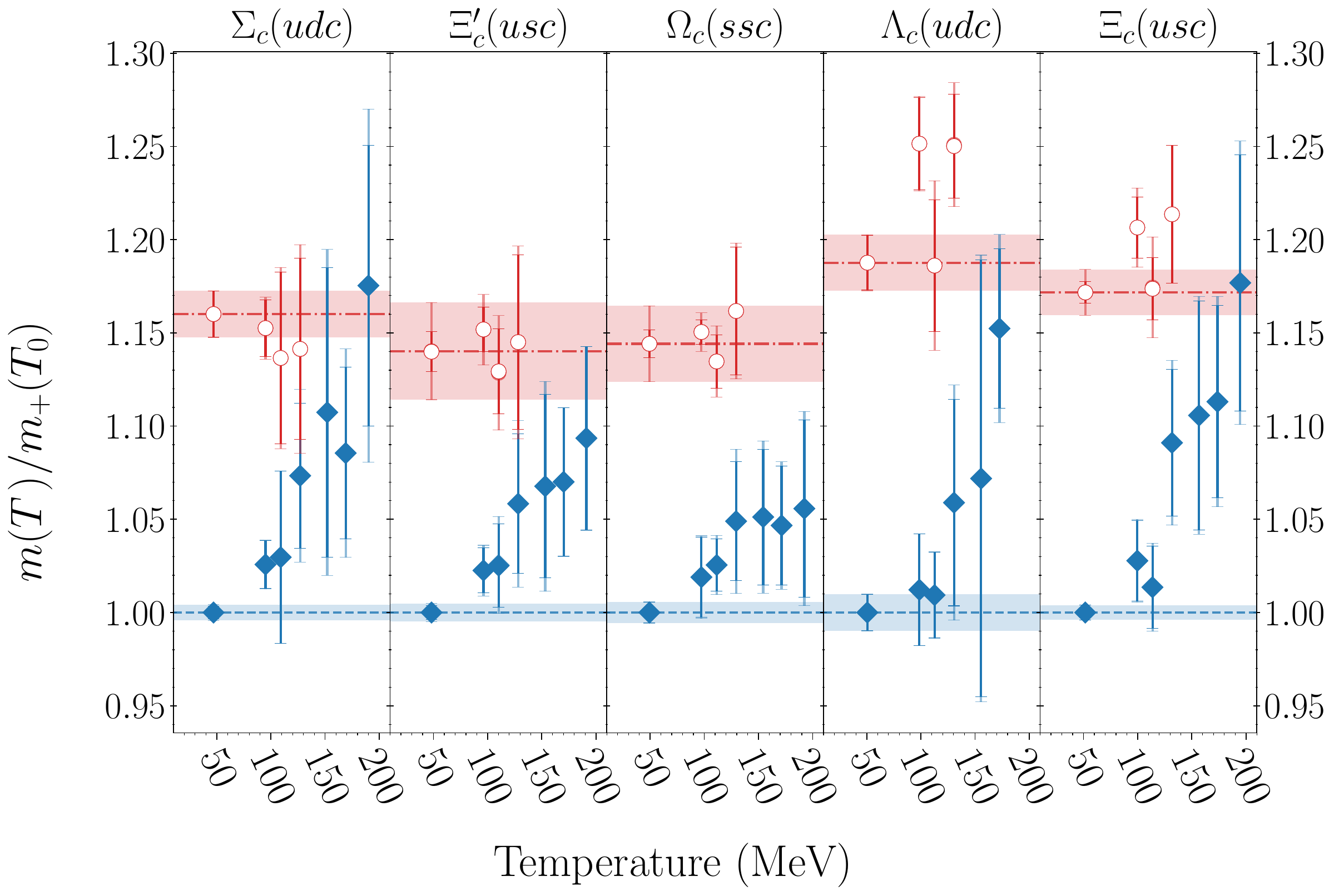}%
  \includegraphics[width=0.5\columnwidth]{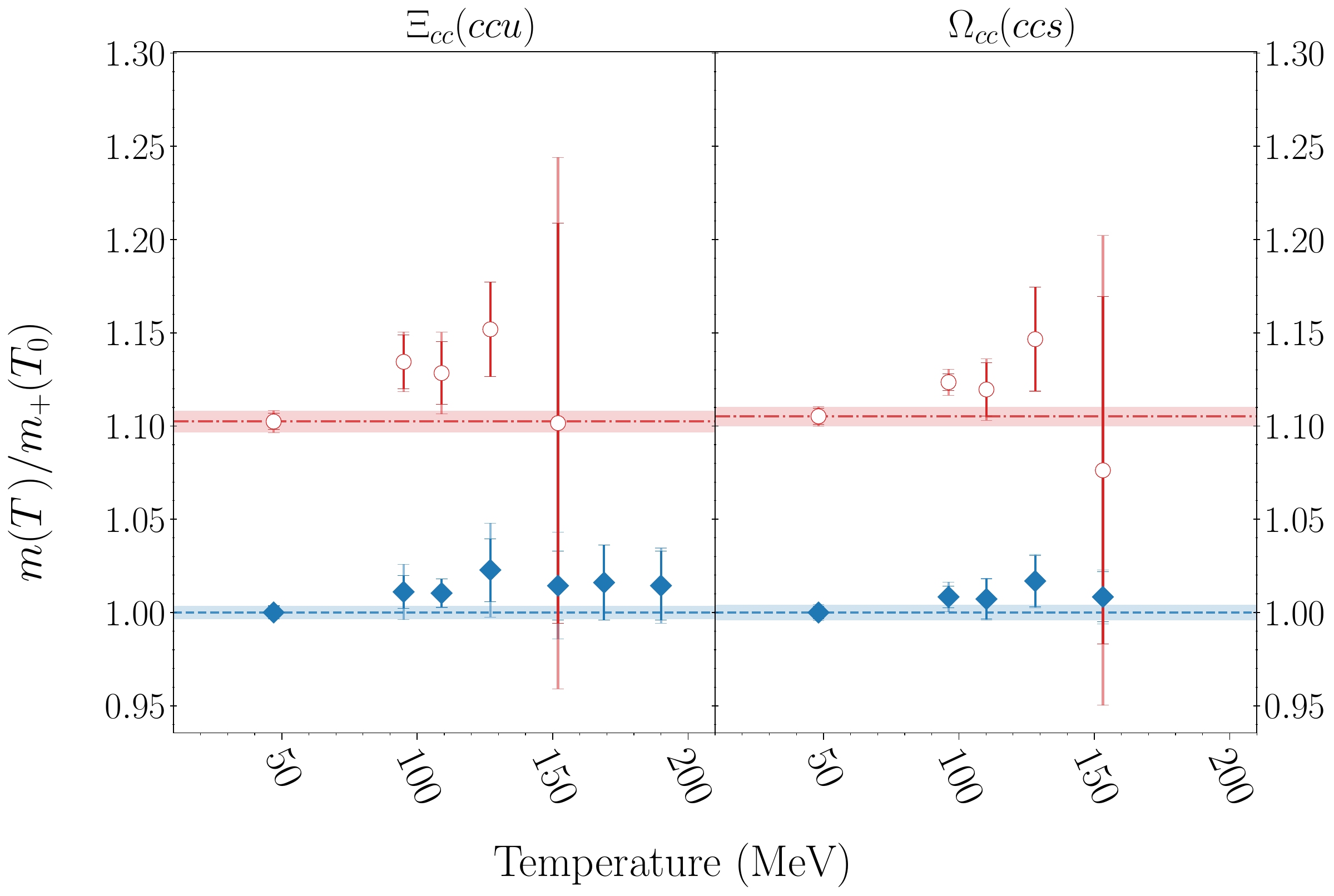}%
  \caption{\label{fig:spectJ1_2P}Ground state masses of singly charmed spin $1/2$ baryons, normalised with the positive parity ground state mass at the lowest temperature $T_0$ (corresponding to $N_\tau =128)$, as a function of temperature. Filled (open) symbols are used for positive (negative) parity states. The inner error bar represents the statistical uncertainty and the outer incorporates the systematic from the choice of averaging method. Horizontal dashed lines show the result from the lowest temperature. The uncertainty of the lowest temperature reflects the relative uncertainty of the lowest temperature mass. Masses are shown only when the ratio analysis suggests the mass can be extracted using an exponential ansatz.
  }
\end{figure}
\par
By studying the double ratio above, the temperatures at which the spectral function of the correlator changes only minimally from the narrowly peaked zero-temperature case are determined and we fit to \eqnr{eqn:barfit} at these temperatures. The resulting masses are shown in \Fig{fig:spectJ1_2P} where two key features emerge. Firstly, the negative parity states are more strongly affected by temperature - as evidenced by the smaller temperature range where we are confident that fits are appropriate compared to the positive parity states, and secondly in that the double charmed baryons $\Xi_{cc}(ccu)$ and $\Omega_{cc}(ccs)$ are much less strongly affected than the singly charmed baryons. In particular the positive parity $\Xi_{cc}(ccu)$ mass remains approximately constant well past $T_c$, up to $T=190$ MeV.
\section{Discussion \& Conclusions}
The temperature dependence of spectral quantities such as the mass and the width are important to our understanding of the formation and evolution of a quark gluon plasma. Here, we have discussed these in the context of the $\Upsilon$ hadrons where we observe little variation as the light quark mass is varied from just lighter than the strange quark to significantly lighter (albeit still heavier than the physical light quark mass). We then compared two analysis methods, used to determine the mass of the state and found that they show remarkable qualitative and semi-quantitative agreement. In particular, a slight negative mass shift as the temperature increases is indicated. This shift is $\sim30\text{--}40$ MeV at $T=250$ MeV -- significantly past the chiral transition temperature, $T_c$.
\par
As charm baryons are sometimes thought to behave similarly to heavy quarkonia, they are also investigated. Here we have carefully examined a \enquote{double ratio} to examine the change in the spectral function from a zero-temperature ground state ansatz. Excited state effects were reduced by the double ratio which assumes that excited states at zero and finite temperature are broadly similar. Following this ratio analysis, we present the mass of the spin $1/2$ charm baryons as a function of temperature in \Fig{fig:spectJ1_2P}. Here the most striking results are the remarkably stable $\Xi_{cc}\rb{ccu}$ (up to $\sim 190$ MeV) and the observation that the negative parity sector is more strongly affected by temperature than the positive across all charm baryon states.
\par
In the future we will calculate bottomonium spectral functions using a simulation with parameters close to Generation 2 but with twice the number of temporal points at each temperature~\cite{Skullerud:2022yjr}. This will be particularly useful for the MEM analysis at higher temperature in order to understand the impact of using only a small number of data points in the spectral function calculation.
\section{Software \& Data}
The software and data for the charm baryon results are available at \Refl{aarts_2023_8275973}. We anticipate making the NRQCD results available after a future publication. Information about the gaugefield ensembles are available from \Refltwo{aarts_2023_8403827}{aarts_2024_10636046}. NRQCD correlators were produced using the package available from \Refl{FASTNRQCD} while the relativistic charm baryon correlators were producing with the \textsc{openQCD} based code available at \Refl{glesaaen_jonas_2018_2217028}. This analysis makes extensive use of the \textsc{python} packages \textsc{gvar}~\cite{peter_lepage_2025_gvar} and \textsc{lsqfit}~\cite{peter_lepage_2025_lsqfit}.

\acknowledgments
We acknowledge EuroHPC Joint Undertaking for awarding the project EHPC-EXT-2023E01-010 access to LUMI-C, Finland. This work used the DiRAC Data Intensive service (DIaL2 \& DIaL) at the University of Leicester, managed by the University of Leicester Research Computing Service on behalf of the STFC DiRAC HPC Facility (www.dirac.ac.uk). The DiRAC service at Leicester was funded by BEIS, UKRI and STFC capital funding and STFC operations grants. This work used the DiRAC Extreme Scaling service (Tesseract) at the University of Edinburgh, managed by the Edinburgh Parallel Computing Centre on behalf of the STFC DiRAC HPC Facility (www.dirac.ac.uk). The DiRAC service at Edinburgh was funded by BEIS, UKRI and STFC capital funding and STFC operations grants. DiRAC is part of the UKRI Digital Research Infrastructure. This work was performed using the PRACE Marconi-KNL resources hosted by CINECA, Italy. We acknowledge the support of the Supercomputing Wales project, which is part-funded by the European Regional Development Fund (ERDF) via Welsh Government. This work is supported by STFC grant ST/X000648/1 and The Royal Society Newton International Fellowship. RB acknowledges support from a Science Foundation Ireland Frontiers for the Future Project award with grant number SFI-21/FFP-P/10186. RHD acknowledges support from Taighde Éireann – Research Ireland under Grant number GOIPG/2024/3507. We are grateful to the Hadron Spectrum Collaboration for the use of their zero-temperature ensemble.

\FloatBarrier
\bibliographystyle{JHEP}
\bibliography{skeleton}

\end{document}